\documentclass{aa}  

\usepackage{graphicx}
\usepackage{txfonts}
\begin{document}

   \title{Destruction of eccentric planetesimals by ram pressure and erosion}

   \author{Tunahan Demirci \and Niclas Schneider \and Jens Teiser \and
          Gerhard Wurm}

   \institute{University of Duisburg-Essen, Faculty of Physics, Lotharstr. 1, 47057 Duisburg, Germany\\
              \email{tunahan.demirci@uni-due.de}
             }

   \date{Received XXX; accepted YYY}

        \abstract{Small, pebble-sized objects and large  bodies of planetesimal size both play important roles in planet formation. They form the evolutionary steps of dust growth in their own respect. However, at later times, they are also thought to provide background populations of mass that larger bodies might feed upon. What we suggest in this work is that starting at times of viscous stirring, planetesimals on eccentric orbits could simply explode as they become supersonic in comparison to small, porous planetary bodies entering Earth's atmosphere. We present a toy model of planetesimal motion and destruction to show the key aspects of this process. The consequences are quite severe. At all times, it is shown that only planetesimals on more or less circular orbits exist in the inner disk. After the destruction of a planetesimal, the remaining matter is continuously redistributed to the pebble reservoir of the protoplanetary disk. Since destruction typically occurs at small stellar distances due to supersonic speeds, it is expected to boost pebble accretion in the inner protoplanetary disk as one of its main effects.} 


   \maketitle
%

\section{Introduction}

This paper is focused on the fragile nature of planetesimals. While the details of each case may vary, planetesimals generally start out as growing dust aggregates in protoplanetary disks \citep{Blum2008, Johansen2014}. It was once believed that dust grows through sticking collisions all the way up to km-size planetesimals. This assumption did not hold up to verification and this is no longer considered a viable path to explaining the formation of the first planetesimals. Bouncing and fragmentation barriers simply prevent infinite collisional growth in such scenarios \citep{Weidling2009, Zsom2010,Beitz2011,Weidling2012, Deckers2013, Deckers2014, Kruss2016,Demirci2017}.

Ignoring these barriers for a moment and assuming planetesimals have formed through sticking collisions, they would be compact dusty objects with tensile strengths of about $1\,\rm kPa$ \citep{Blum2006, Katsuragi2017}. To grasp the overall range of tensile strengths to be expected for small bodies, this might be compared to meteorites.
Meteorites are very compact in comparison to these dust agglomerates. Such a compression grade could be achieved through high-velocity impacts and, in any case, the tensile strength is greater up to a few $10\, \mathrm{kPa}$ \citep{Sebastian2020}. Such values are also found for cometary meteorides, which ultimately disintegrate in Earth's upper atmosphere \citep{TrigoRodriguez2006}.  These values are still small compared to those of solid rocks, for which it takes 20 MPa to GPa to be crushed \citep{Cohn1981, Leinhardt2009,Flynn2018,Borovicka2020}. Rigid ice-dust mixtures are also in the range of 1-20 MPa \citep{Lange1983}. Taking these bodies into consideration, hydrodynamic destruction in protoplanetary disks is not the first rationale that comes to mind.

Nonetheless, to stand in the way of any obstacles with regard to planetesimal growth (bouncing or fragmentation), recent models have considered the notion that various drag instabilities caused by the interaction between pebble size dust aggregates and the disc's gas provide enough particle concentration to make the jump right away from pebbles to planetesimals in a final gravitational collapse \citep{Youdin2005, Johansen2007, Johansen2015, Krapp2020}. This collapse is still gentle enough so that most of the dust aggregates are not compressed further but pebble-pile planetesimals can still form \citep{Lambrechts2012, Johansen2014, Blum2017, Johansen2017, Bitsch2018, Bitsch2019}. In this case, compared to the $1\,\rm kPa$ tensile strength for compact dust, only tensile strengths of about $1\,\rm Pa$ result \citep{Skorov2012}.

Planetesimals on a circular orbit in the sub-Keplerian disk encounter a headwind of about $50\, \mathrm{m}\,\mathrm{s}^{-1}$ \citep{Weidenschilling1993}. This is already enough to erode planetesimals, even though the pressure is low \citep{Demirci2019, Demirci2020, Kruss2020, Rozner2020, Grishin2020}, while making note that the needed shear stress at the surface can be orders of magnitude lower than the tensile strength \citep{Shao2000}. Erosion is already known to occur on simple circular orbits  in the inner $1\,\rm AU$  \citep{Demirci2019,Demirci2020}. If the planetesimal’s orbit is becoming eccentric, the relative speed between the planetesimal and gas increases. This extends the range of conditions under which planetesimals are unstable \citep{Schaffer2020}. Even compact dusty bodies can get eroded on slightly eccentric orbits \citep{Paraskov2006}.
The question is, however, how far eccentricity can account for these effects before this problem is shown to not only affect some planetesimals close to the star, but a wide range of stellar distances as well.

As seen below, only small eccentricities are needed for the planetesimal to reach a supersonic velocity relative to gas. It is known based on the physics of aircraft travel that the drag force strongly increases as the aircraft moves into the transonic regime, where only a part of the stream around a body is supersonic. Besides this, we must also consider the forces that impact bodies  moving at a supersonic speed that is several times the speed of sound.

Laboratory experiments at low pressure with dust aggregates are not easily carried out in this regime, however, nature provides insights that demonstrate that such settings are not connected to planetesimal survival. Small, present-day bodies of the solar system regularly impact Earth. The most porous of them never reach the ground. One of the most impressive examples might be the Tunguska event of 1908,
when an object that was likely to be of cometary origins exploded in the atmosphere, devastating $2000\,\mathrm{km}^2$ of ground in Siberia.

A smaller, more recent event of
some public recognition involved the Chelyabinsk meteor, which exploded in the atmosphere, with its remnants now characterized as LL5 chondrite \citep{Galimov2013}. In contrast to cometary nuclei, chondrites are rocky bodies. These chondrites have their origin in the asteroid belt and they have already solidified quite significantly compared to planetesimals that have formed from pebbles or dust. Taking a brief look at heights of about 50 to 100 km, at which meteors typically dissolve, gives us an associated atmospheric pressure of about $1\,\rm Pa$ to $100\,\rm Pa$ \citep{Molau2008}. This is comparable to the pressure range in the inner solar nebula \citep{Hayashi1981}.

These bodies enter the atmosphere at speeds of tens of $\mathrm{km}\,\mathrm{s}^{-1},$ or a few tens of  the speed of sound, which typically  is faster than the relative velocities in a protoplanetary disk even for eccentric orbits. However, it still suggests that much more weakly bound, supersonic planetesimals are just unstable in a protoplanetary disk.

If these bodies are destroyed, it carries several implications. We note here that this is  a brief work based on a toy model, so we do not follow up on these implications in detail, but the matter concerns many aspects of planet formation, some of which we mention in the following. The instantaneous destruction of planetesimals on eccentric orbits might be a severe problem for shock models for chondrule formation, which are based on supersonic planetesimals \citep{Ciesla2004, Boss2005, Cuzzi2006} and which might not regularly exist. Also, the dynamics of runaway growth of pebbles and planetesimal accretion by larger objects will change severely if only planetesimals on circular orbits are expected to survive \citep{Ormel2010, Lambrechts2012, Johansen2017, Bitsch2018, Bitsch2019, Liu2019,Liu2019b, Bruegger2020}. Pebble accretion by large bodies might benefit, as more pebbles might be available after the destruction of planetesimals on eccentric orbits. Also, the location of planet formation might shift, as the destruction of an eccentric planetesimal dumps its material closer to the periastron of the respective orbit. This could even prove beneficial with regard to the process of planet formation \citep{Drazkowska2016}.

\section{Conditions for destruction}
\label{sec:cond}
\begin{figure}
        \centering
        \includegraphics[width=0.9 \columnwidth]{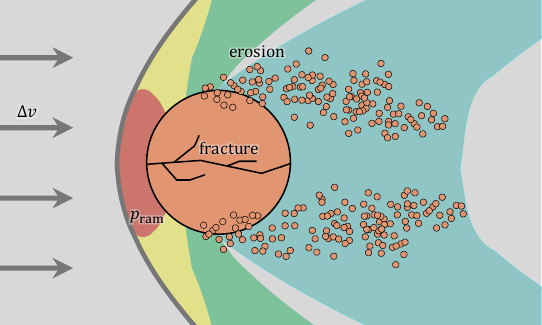}
        \caption{Scheme of a planetesimal with supersonic relative velocity and indicated fracture and erosion as the main destruction mechanisms.}
        \label{fig:sketch}
\end{figure}

We identify two main mechanisms for planetesimal destruction: erosion and fracturing (Fig. \ref{fig:sketch}). Erosion starts when the wall shear stress at the surface, $\tau_\mathrm{wall},$ is greater than the critical shear stress for the specific planetesimal, $\tau_{\rm erosion},$ \citep{Shao2000, Demirci2019,Demirci2020, Grishin2020, Rozner2020, Schaffer2020}:
\begin{equation}
\label{eq:erosioncond}
\frac{\tau_{\rm wall}}{\tau_{\rm erosion}} > 1.
\end{equation}

The wall shear stress depends on the gas density, $\rho$, the dynamic viscosity, $\mu,$ and, furthermore, on the relative velocity between gas and planetesimal, $\Delta v,$ and the radius of the planetesimal, $R_{\rm P}$,
\begin{equation}
\tau_\mathrm{wall}=\frac{1}{5}\left(\frac{\rho~ \mu ~\Delta v^3}{R_\mathrm{P}} \right)^{\frac{1}{2}}.
\end{equation}

The critical threshold for erosion was determined experimentally by \cite{Demirci2019,Demirci2020} and depends on the particle density, $\rho_{\rm P}$, the gravitational acceleration on the surface of the planetesimal, $g_{\rm P},$ and the size, $d_{\rm particle},$ of the pebbles that make up the planetesimal and the effective surface energy, $\gamma_{\rm eff}$. Additionally, the Cunningham correction for slip-flow, $f_\mathrm{C}(\beta^{-1} \mathrm{Kn}),$ is applied. The term $\beta^{-1} \mathrm{Kn}$ gives an effective Knudsen number, accounting for the different length scales of a pebble-pile planetesimal. The coefficient $\alpha$ describes the efficiency of the drag in lifting particles from the surface (see \citet{Demirci2020} for details).

\begin{equation}
\tau_\mathrm{erosion}=\alpha f_\mathrm{C}(\beta^{-1} \mathrm{Kn})\left( \frac{\gamma_\mathrm{eff}}{d_\mathrm{particle}} + \frac{1}{9} \rho_\mathrm{P} g_\mathrm{P} d_\mathrm{particle} \right).
\end{equation}

Beside the gas flow on the surface of the planetesimal, a dynamic pressure $p_{\rm ram}$ acts on the planetesimal:
\begin{equation}
\label{eq:pRAM}
p_{\rm ram} = \dfrac{1}{2} \rho \Delta \! v ^2.
\end{equation}
A planetesimal is disrupted completely when the dynamic pressure exceeds the tensile strength $p_{\rm TS}$ \citep{Valletta2019}.
The condition for planetesimal fragmentation is:
\begin{equation}
\label{eq:tensilestr}
\frac{p_{\rm ram}}{p_{\rm TS}}>1.
\end{equation}

\section{Planetesimals on eccentric orbits}
\label{sec:eccentric}

\begin{figure}
        \centering
        \includegraphics[width=1\columnwidth]{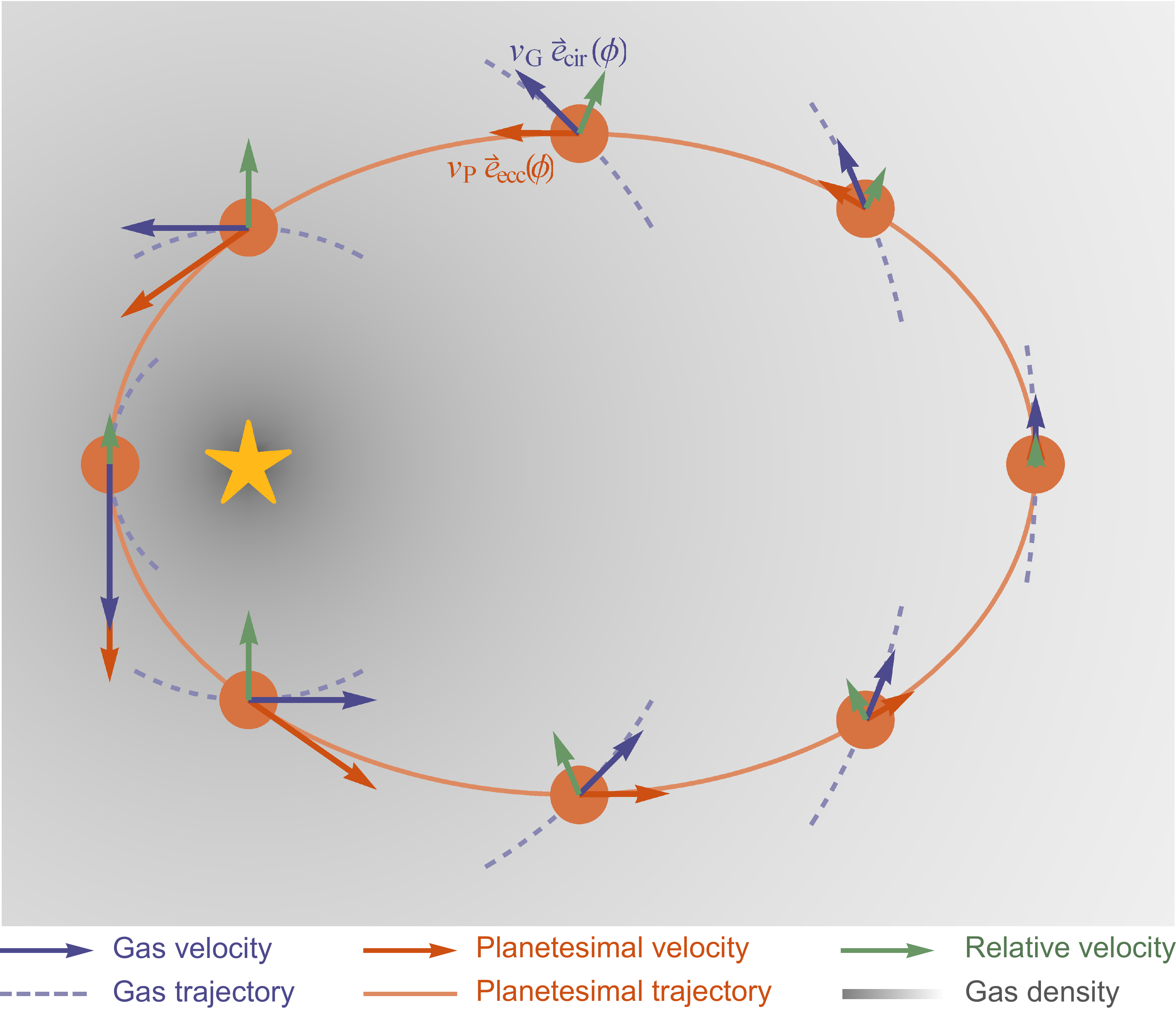}
        \caption{Position of the planetesimal and the corresponding local velocities of the gas, the planetesimal, and the relative velocity. The gray background indicates the gas density changing with the distance to the star.}
        \label{fig:sketch}
\end{figure}

A recent work by \citet{Mai2020} treats a somewhat related problem in their discussion of the stripping of a protoplanetary atmosphere on eccentric orbits.
The eccentric orbit of a planetesimal around the central star is described using the phase-dependent radial distance,
\begin{equation}
\label{eq:r}
r(a,e,\phi)=a \frac{1-e^2}{1-e \cos{\phi}},
\end{equation}
with the semimajor axis, $a$, the eccentricity, $e,$ and phase angle, $\phi \in [0,2\pi]$. The angle, $\phi=0=2\pi,$ describes the position at the apastron of the orbit and $\phi=\pi$ describes the position at the periastron. As planetesimals are assumed to form in the mid-plane \citep{Johansen2014}, we are looking at a two-dimensional problem, where the inclination is set to $i=0$. In the case of an inclined orbit, the relative velocities between gas and planetesimal will further increase, so the restriction to a two-dimensional geometry is rather conservative.

The velocity of the planetesimal on the eccentric orbit depends on the radial distance, $r,$ to the star and is expressed with the Vis-viva equation:
\begin{equation}
\label{eq:v_per}
\vec{v}_{\rm P} = v_{\rm K}(r)  \sqrt{2-\frac{r}{a}} ~\vec{e}_\mathrm{ecc}(\phi),
\end{equation}
where $v_{\rm K}(r)=\sqrt{\frac{G M_\odot}{r}}$ is the Kepler velocity on a circular orbit. In contrast to the planetesimal, the gas of the protoplanetary disk moves on circular orbits with the velocity,
\begin{equation}
\label{eq:v_G}
\vec{v}_{\rm G} =  v_{\rm K}(r)  \sqrt{1-\eta(r)}~\vec{e}_\mathrm{cir}(\phi).
\end{equation}
The term $\sqrt{1-\eta(r)}$ results from the pressure gradient within the protoplanetary disk. The unit vectors, $\vec{e}_\mathrm{cir}(\phi)$ and $\vec{e}_\mathrm{ecc}(\phi),$ show the tangential directions of the circular and eccentric orbit, respectively (see Fig. \ref{fig:sketch}). The relative velocity between planetesimal and gas is:\ 
\begin{equation}
\label{eq:velocity}
\Delta \! v = \| \vec{v}_{\rm P} - \vec{v}_{\rm G} \|. 
\end{equation}

The relative velocity strongly depends on the orbit parameters ($a,e,\phi$) of the planetesimal. Both destruction mechanisms described in this work - erosion and fracturing - depend on the relative velocity, $\Delta \! v,$ and the gas density, $\rho$. The gas density in the protoplanetary disk depends on the radial distance to the star and is expressed for the minimum-mass solar nebula \citep{Hayashi1981} as:\ 

\begin{equation}
\label{eq:rho}
\rho(r)=\rho_0 \cdot \left(r/1\,\mathrm{AU}\right)^{-11/4}
,\end{equation}
with $\rho_0=1.4\cdot 10^{-6}\,\mathrm{kg}\,\mathrm{m}^{-3}$ being the gas density at $1\,\mathrm{AU}$.

\begin{figure}
        \centering
        \includegraphics[width=\columnwidth]{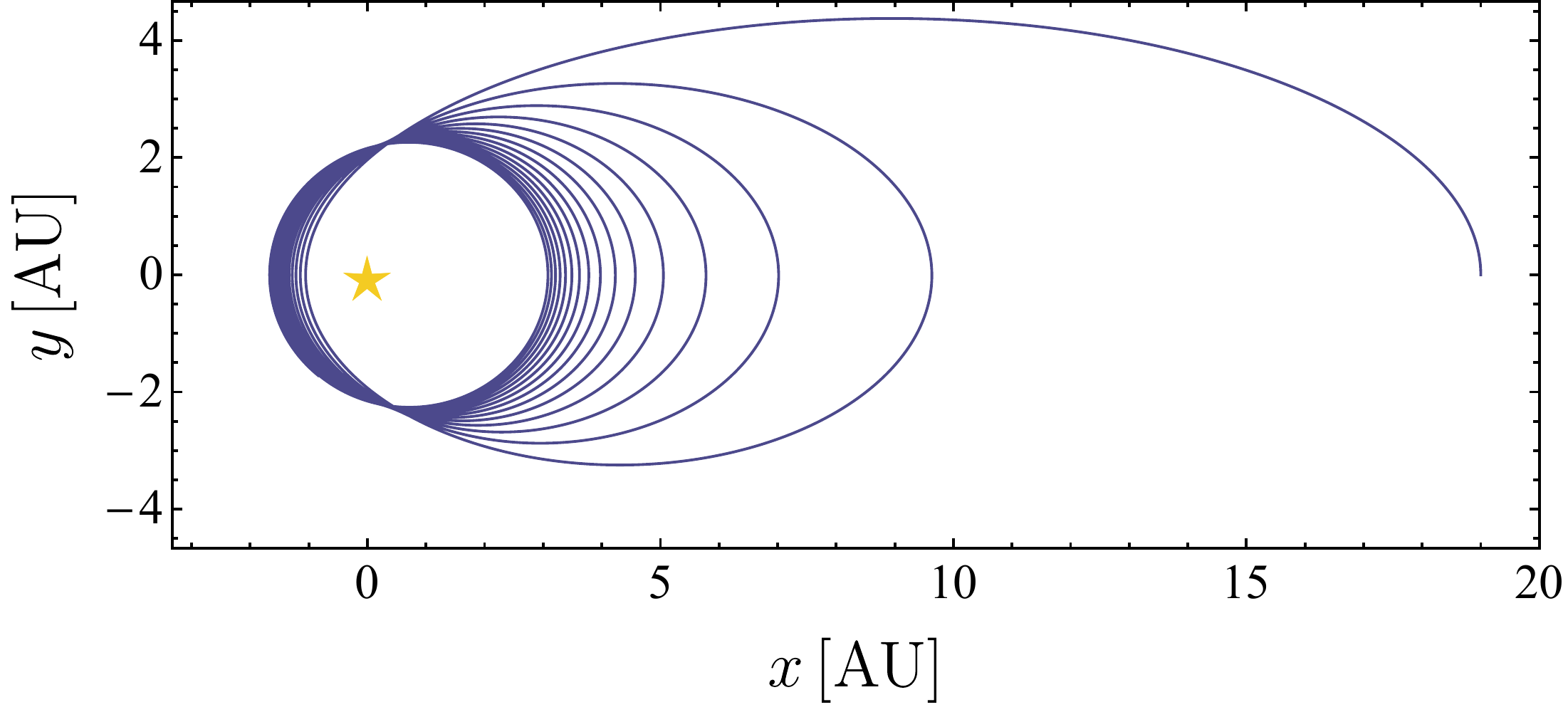}
        \caption{$1\,\mathrm{km}$-sized planetesimal starting at the apastron with $a=10\,\mathrm{AU}$ and $e=0.9$ loses angular momentum with every orbit around the star due to gas drag.}
        \label{fig:ecc_traj}
\end{figure}

 Planetesimals on eccentric orbits are particularly affected by gas drag due to their relative velocity to the gas. Therefore, they lose angular momentum and eccentricity declines over time. In Fig. \ref{fig:ecc_traj}, we show an exemplary trajectory of a planetesimal with high eccentricity. Due to the high relative velocities and the resulting Reynolds numbers, the Newtonian drag law applies.

\begin{figure}
        \centering
        \includegraphics[width=0.9\columnwidth]{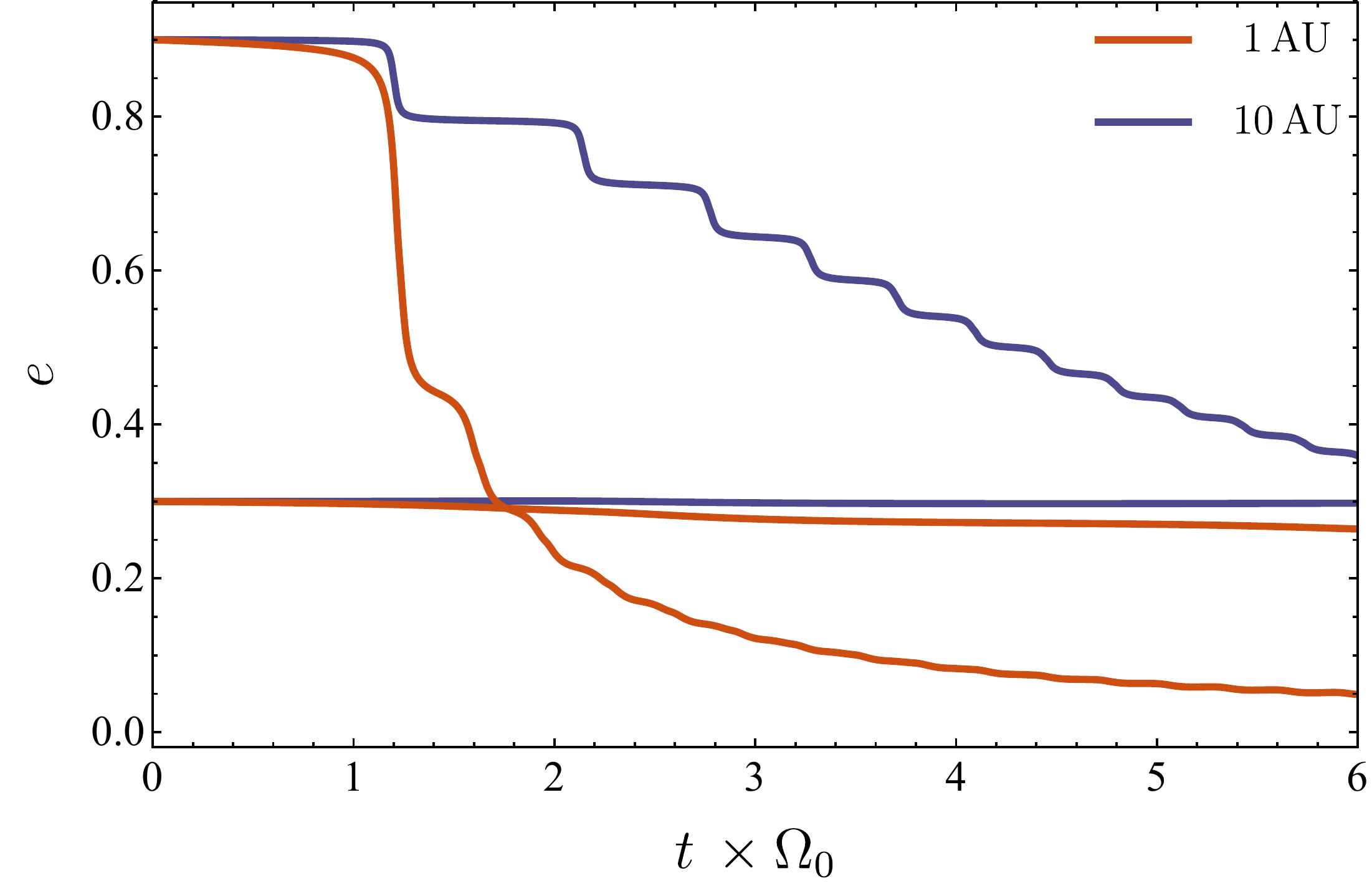}
        \caption{Eccentricity in the dependence of the time multiplied with the initial Kepler frequency $\Omega_0=\sqrt{G M_\odot r^{-3}}$ for $1\,\mathrm{km}$-sized planetesimals. The planetesimals have an initial axis of 1 AU (red lines) or 10 AU (blue lines) and an initial eccentricity of 0.3 or 0.9.}
        \label{fig:ecc_loss}
\end{figure}

The eccentricity of different planetesimals over time is shown in Fig. \ref{fig:ecc_loss}. The smaller the semimajor axis, the more the planetesimals are affected by the gas drag and the faster they lose eccentricity. In addition, high initial eccentricities decline on shorter timescales.

This evolution should be kept in mind for detailed calculations. Here, we only give a first estimate on the stability and we note that fracturing and erosion thresholds presented in the next section (Sect. \ref{sec:results}) are a worst-case scenario and do not present a time-dependent evolution of eccentric planetesimals.

\section{Results}
\label{sec:results}

\subsection{Fracturing}

\begin{figure}
        \centering
        \includegraphics[width=\columnwidth]{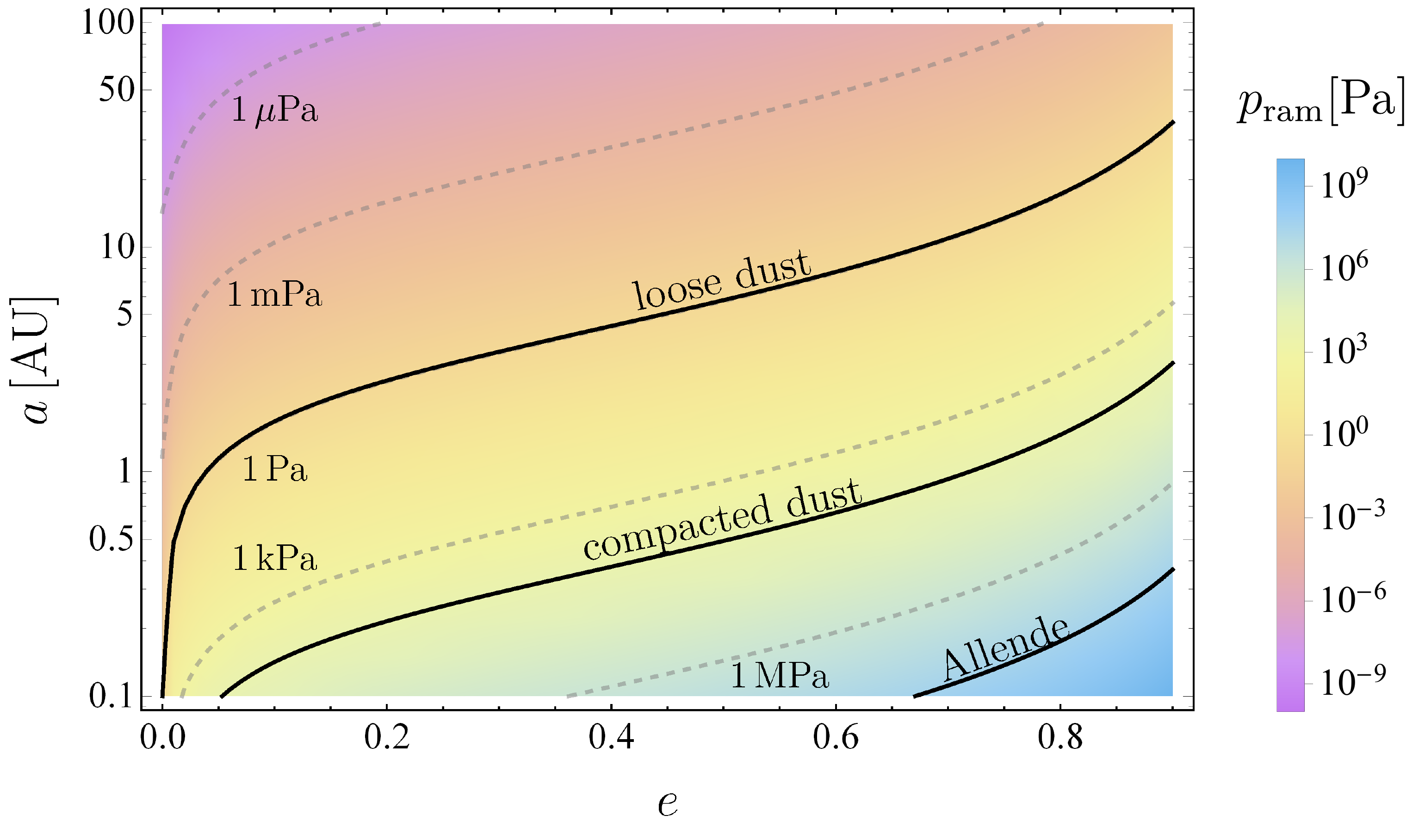}
        \caption{Maximum ram pressure $p_\mathrm{ram}$ in dependence of the eccentricity $e$ and the semimajor axis $a$. For comparison, the tensile strengths of a loose \citep{Skorov2012} and a compacted dust sample \citep{Steinpilz2019} are shown as well as the tensile strength of the Allende meteorite \citep{Flynn2018}. The planetesimal is destroyed by the gas drag, if the ram pressure exceeds the tensile strength of the planetesimal material (below the corresponding curve).}
        \label{fig:ram}
\end{figure}

\begin{figure}
        \centering
        \includegraphics[width=\columnwidth]{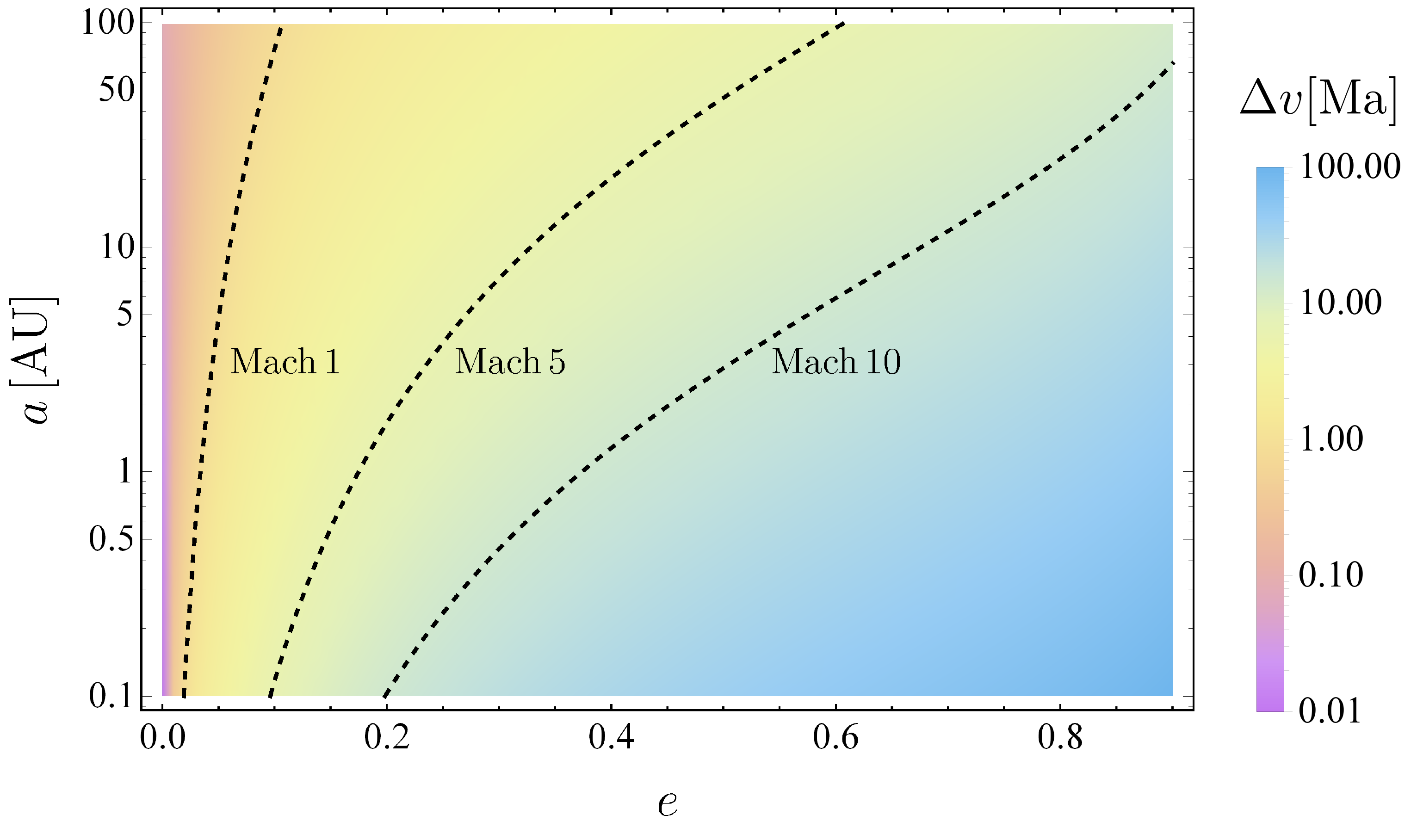}
        \caption{Relative velocity $\Delta \! v$ in units of the local sound speed for the maximum ram pressure in dependence of the eccentricity $e$ and semimajor axis $a$. Even for small eccentricities, the relative velocity becomes supersonic.}
        \label{fig:velocity}
\end{figure}

In Fig. \ref{fig:ram}, we plotted the maximum ram pressure for a wide range of orbit parameters $a \in [0.1\,\mathrm{AU}, 100\,\mathrm{AU}]$ and $e \in [0,0.9]$. We illustrate the maximum ram pressure is achieved at a phase angle of about $\phi \approx \pi/2$.  For comparison, the tensile strengths of a loose \citep{Skorov2012} and compacted dust sample \citep{Steinpilz2019} and of the Allende meteorite \citep{Flynn2018}. The planetesimal is destroyed by the gas drag if the ram pressure exceeds the tensile strength of the planetesimal material. For meteorite-like bodies, this is only the case on highly eccentric orbits very close to the star. Planetesimals with a tensile strength comparable to that of loose dust are prone to destruction at further distances to the star, on orbits with a much smaller eccentricity. A planetesimal with a semimajor axis of  $1\mathrm{AU}$ with that low tensile strength would be destroyed by the ram pressure if the eccentricity only exceeds $e>0.03$. A planetesimal with a tensile strength of $10\,\mathrm{kPa}$, which is comparable to the strength of compacted dust \citep{Steinpilz2019}, is stable against disruption in a wider parameter space. For this kind of planetesimal with a semimajor axis of $1\mathrm{AU,}$ an eccentricity of $e>0.7$ is needed to destroy it. 

Figure \ref{fig:velocity} shows the relative velocity, $\Delta \! v,$ in units of the local sound speed for the maximum ram pressure in dependence of the eccentricity, $e,$ and the semimajor axis, $a$. 
Even if the threshold for fracturing is exceeded, this does not necessarily mean that the planetesimal is destroyed. \cite{Pollack1979} argued that as long as the gravitational force of the planetesimal is larger than the drag force acting on a fragment, the fragments will stay together. However, this is strongly dependent on the size distribution of the outcome of the fragmentation, on the drag force, and the remaining self-gravity of the planetesimal.
It must be noted that even for small eccentricities, the relative velocity becomes supersonic and, therefore, the gas can no longer be treated as non-compressible. Consequently, the considered ram pressure is a minimum estimation of the acting pressure onto the planetesimal, as in the transonic and supersonic regimes additional forces act due to the resulting shock waves. We neglect these effects for simplicity, but we do note that the used ram pressure is only a lower limit.

\subsection{Erosion}

\begin{figure}
        \centering
        \includegraphics[width=\columnwidth]{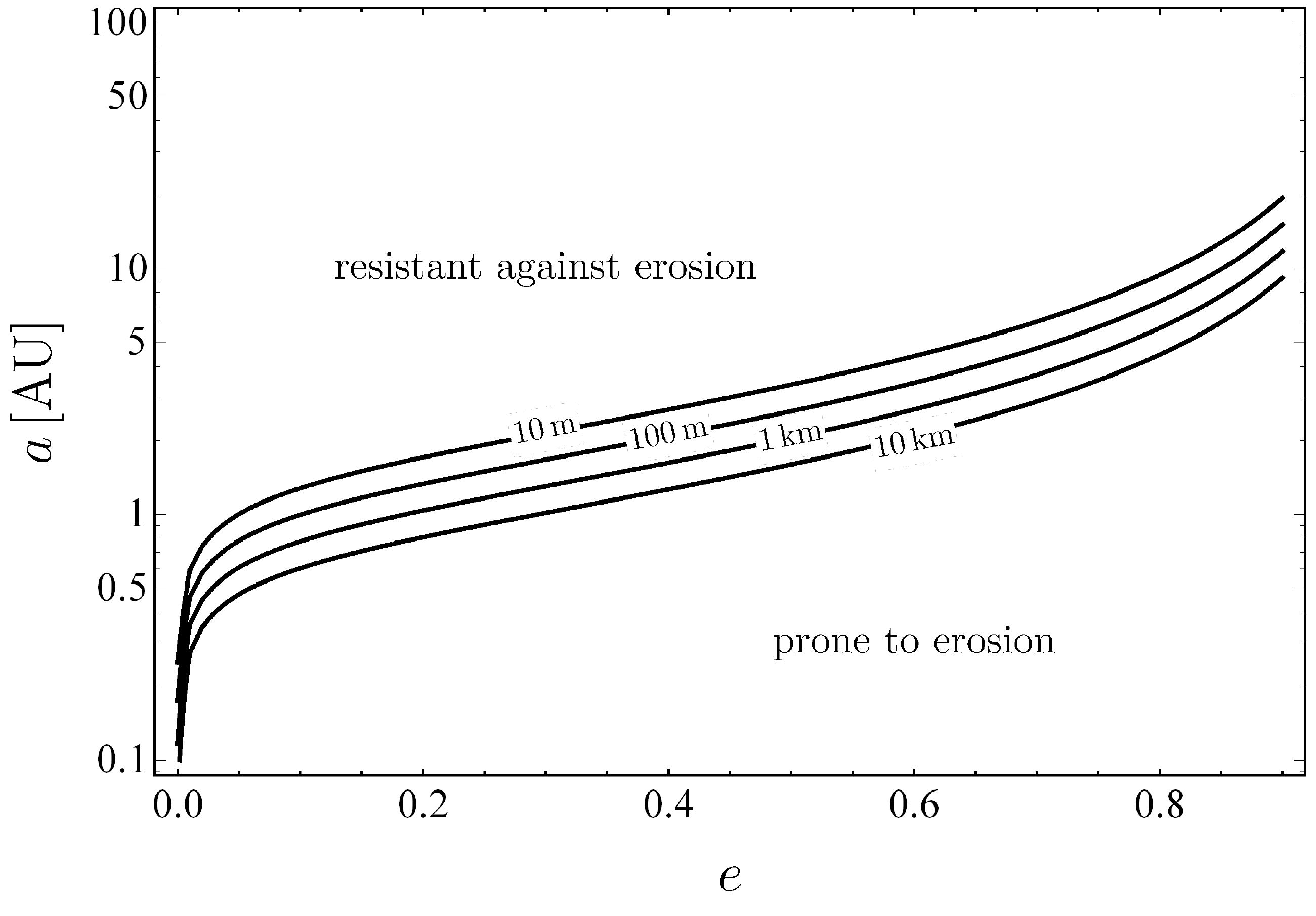}
        \caption{Erosion thresholds for a loosely bound pebble-pile planetesimal with radius, $R_\mathrm{P},$ in dependence of the eccentricity, $e,$ and semimajor axis, $a$. The planetesimals consist of millimetre-sized pebbles and the cohesion between the pebbles is described with the effective surface energy, $\gamma_\mathrm{eff}=10^{-4}\,\mathrm{N}\,\mathrm{m}^{-1}$ taken from experiments \citep{Demirci2020}.}
        \label{fig:erosion}
\end{figure}

If the planetesimal is a loose collection of millimeter sized pebbles (a so-called pebble-pile planetesimal) then it is also prone to destruction by erosion. We plotted for the same orbital parameter range ($a,e$), the erosion threshold of planetesimals of different sizes $R_\mathrm{P}=\{10\,\mathrm{m},100\,\mathrm{m},1\,\mathrm{km},10\,\mathrm{km}\}$ (see Fig. \ref{fig:erosion}). The planetesimals consist of millimetre-sized pebbles and the cohesion between the pebbles is described with the effective surface energy of $\gamma_\mathrm{eff}=10^{-4}\,\mathrm{N}\,\mathrm{m}^{-1},$ taken from experiments \citep{Demirci2020}. The illustrated lines in Fig. \ref{fig:erosion} describe the orbital parameters ($a,e$) at the erosion threshold ($\tau_{\rm wall}=\tau_{\rm erosion}$). \citet{Demirci2019} showed that millimetre-sized pebbles which are lifted by the wind will overcome the self-gravity of a $10\,\mathrm{km}$-sized planetesimal on a circular orbit at $1\,\mathrm{AU}$ and leave for good, so that the erosion of planetesimals studied in this work (higher wind velocities and smaller planetesimals) will always result in complete destruction. The bigger a planetesimal is, the more stable it is against erosion by protoplanetary winds. This is due to the increasing gravitational acceleration, $g_\mathrm{P}$, and the decreasing wall shear stress, $\tau_\mathrm{wall},$ for increasing planetesimal radii, $R_\mathrm{P}$. For example, a planetesimal with an eccentricity $e=0.1$ and a radius $R_\mathrm{P}=10\,\mathrm{m}$ is destroyed at $a\lesssim 1.3\,\mathrm{AU}$, whereas a planetesimal with  $R_\mathrm{P}=10\,\mathrm{km}$ is only eroded at $a\lesssim 0.6\,\mathrm{AU}$. Compared with the destruction due to ram pressure, the threshold for wind erosion falls between the fracturing threshold of loose and compacted dust for most of the parameter space ($a,e$) but small eccentricities (cf. Fig. \ref{fig:ram}). There is a crossover where both effects are equal due to the different dependence on the relative velocity between planetesimal and gas, which depends on the eccentricity. Therefore, in general, fracturing is the dominant process for eccentric orbits, however, small planetesimals on circular orbits or for very small $e$ are destroyed by erosion (cf. Figs. \ref{fig:ram} and \ref{fig:erosion}).

\section{Conclusion}

Pebbles and planetesimals on eccentric orbits have a high velocity relative to the gas of the protoplanetary disk. Their high eccentricity can lead to the destruction of the protoplanetary body. The two destruction mechanisms discussed in this work are fracturing and erosion, both of which are triggered at high velocities and high gas densities. We calculated the stability regions of these planetary objects to find that only planetesimals with very small eccentricities can survive the destruction mechanisms underway in the inner protoplanetary disk. Close to the star, fracturing and erosion are processes that eliminate planetesimals on eccentric orbits and contribute new pebbles to the protoplanetary disk.

\begin{acknowledgements}
This project is funded by DLR space administration with funds provided by the BMWi under grant 50 WM 1760. N.S. is supported by DFG under grant number WU 321/16-1. We appreciate the constructive review by the anonymous referee.
\end{acknowledgements}

\bibliographystyle{aa}
\bibliography{example}

\end{document}